\documentclass[twocolumn,a4paper,10pt]{article}
\usepackage{amsmath,amsfonts}
\usepackage{epsfig}
\usepackage{times}
\usepackage{epsfig}
\usepackage{subcaption}
\usepackage{float}
\usepackage{wrapfig}
\usepackage{biblatex} 
\usepackage{dirtytalk}
\usepackage{xcolor}
\makeatletter

\newcounter{numbersec}
\renewcommand{\section}[1]{\par\noindent\stepcounter{numbersec}
\par
\vspace{6pt}
\noindent\textbf{\large   \arabic{numbersec} \hspace*{0.3cm} #1 }
\par
\vspace{2pt}
}
\renewcommand{\subsection}[1]{
\par
\vspace{6pt}
\noindent\textbf{#1}
\par
}
\renewcommand{\subsubsection}[1]{%
\par
\vspace{6pt}
\textbf{#1.}
}

\newcommand{\Abstract}{\par\vspace{6pt}\noindent\textbf{\large Abstract}\par\vspace{2pt}}

\newenvironment{References}{
\par\vspace{6pt}\noindent\textbf{\large References}\par\vspace{2pt}
\begin{small}\begin{list}{ }{
\itemsep0mm \parsep0mm\labelsep0mm\leftmargin0mm
}}
{\end{list}\end{small}}

\makeatother

\title{\vspace*{-12mm}
\LARGE \sc \textbf{  
Tripping and laminar--turbulent transition: \\
Implementation in RANS-EVM
}}
\author{ \large \bf \textit{ 
N. Tabatabaei$^{1,2*}$, G. Fahland$^{3}$, A. Stroh$^{3}$, D. Gatti$^{3}$, B. Frohnapfel$^{3}$} \\
\large \bf \textit{M. Atzori$^{1,2}$, R. Vinuesa$^{1,2}$, and P. Schlatter$^{1,2}$}  \\ \\
\bf  $^{1}$ \textit{ SimEx/FLOW, Engineering Mechanics, KTH Royal Institute of Technology, Stockholm, Sweden} \\
\bf  $^{2}$ \textit{Swedish e-Science Research Centre (SeRC), Stockholm, Sweden} \\
\bf  $^{3}$ \textit{ Institute of Fluid Mechanics, Karlsruhe Institute of Technology (KIT), Karlsruhe, Germany} \\ \\
\underline{\bf nargest@kth.se}
}
\date{}

\begin{document}

%


%

\maketitle
\thispagestyle{empty}



%
%

\Abstract
Fundamental fluid-mechanics studies and many engineering developments are based on tripped cases. Therefore, it is essential for CFD simulations to replicate the same forced transition in spite of the availability of advanced transition modelling. In the last decade, both direct and large-eddy simulations (DNS and LES) include tripping methods in an effort to avoid the need for modeling the complex mechanisms associated with the natural transition process, which we would like to bring over to Reynolds-averaged Navier-–Stokes (RANS) turbulence models. 
This paper investigates the necessity and applications of numerical tripping, despite of the  developments in numerical modeling of natural transition. The second goal of this paper is to assess a technique to implement tripping in eddy-viscosity models (EVM) for RANS. A recent approach of turbulence generation, denoted as turbulence-injection method ($k$I), is evaluated and investigated through different test cases ranging from a turbulent boundary layer on a flat plate to the three-dimensional (3D) flow over a wing section. The desired tripping is achieved at the target location and the simulation results compare favourably with the reference results (DNS, LES and measured data). With the application of the model, the challenging transition region can be minimised in a simulation, and consequently more reliable results are obtained. 

\vspace{-6pt}
\section{Introduction}
Modeling the laminar--turbulent transition is still a challenging subject, especially for engineering computational fluid dynamics (CFD). The exact placement of the laminar--turbulent transition has a significant effect on relevant characteristics of the boundary layer and aerodynamics, such as drag, heat transfer and flow separation on \textit{e.g.}\ wings and turbine blades. For instance, the inaccuracy in prediction of the transition onset can result in larger separation regions near the wing trailing edge. Such limitations of CFD simulations increase the discrepancy between experimental and numerical data in the design processes. 

Tripping, which fixes the transition position, has been implemented in wind-tunnel experiments to promote early transition to turbulence in the boundary layer for the past 70 years, because it makes the transition independent of the local condition of the free-stream. In the first part of the paper in section~2, the applications of the tripping technique are discussed, in order to describe why one needs a tripping model rather than a model to simulate  different transition mechanisms. To bring tripping to practical applications, there is a demand to assess the implementation of tripping mechanisms with Reynolds-averaged Navier–Stokes (RANS) approach which can also serve as a design tool. The laminar--turbulent transition models are discussed in section~3, focusing on RANS-EVM. 

With the goal of replicating the same transition point as in the experiments (and corresponding resolved simulations), and removing the uncertainty in the numerical model of forced transition, this study investigates a numerical approach that mimics the effect of a turbulence trip. Tripping in experiments is not always implemented at the natural transition point, rather the target point of forced transition may shift for different reasons such as control. Therefore, a flexible numerical approach is required to control the flow condition, as it has already been investigated in LES and DNS. The present work considers a method to implement tripping with RANS as well as its assessment compared to wind-tunnel experiments, beside the similar tripping approaches in LES and DNS. The methodology and the results are described in sections~3 and 4, respectively. The ultimate goal of the project is to develop a numerical--simulation approach which can represent the complex experimental setup and measurements in a typical wind tunnel, reduce the uncertainty in design of a setup, and thus increase the fidelity of a campaign. This motivates to include three-dimensional (3D) setups as a part of this research, such as complete wind-tunnel setups, in the framework of a \emph{virtual wind tunnel} \cite{insert_1}.

\vspace{-1pt}
\section{Tripping applications and models}\label{sec2}

Due to the variety of transition types and the complexity of modeling the different mechanisms in numerical simulations, one alternative is to start with a correct boundary layer at a certain Reynolds number ($Re_ \theta$), using fully turbulent boundary-layer data as inflow, e.g. from a direct numerical simulation, DNS. In this way, there is no need to go through any transition model, \emph{i.e.}\ the flow is turbulent throughout the computational domain. However, such data are not always available. Another possibility is to skip the whole transition process by tripping the boundary layer. In this way, transition is forced at a prescribed and meaningful location, rather than the natural transition process.

Additionally, the action of forcing transition from a laminar to a turbulent boundary layer (BL) is common in wind-tunnel testing to eliminate the later transition caused by testing at reduced Reynolds numbers ($Re$)~\cite{scale}. For example, at low $Re$ and for an airfoil at stall-angle, it is  essential to ensure that the transition occurs before the laminar flow separation. BL trips are traditionally also used in scale-model testing to aid in scaling the flow characteristics, where duplicating full-scale $Re$s is not feasible in the wind tunnel which leads to the fact that the developed BLs on wind-tunnel models do not correspond to the BLs which develop on full-scale vehicles. For this purpose, tripping devices are placed on models to hasten the BL transition from laminar to turbulent flow. In this way, the characteristics which are sensitive to the condition of the BL are more accurately simulated in tripped cases~\cite{whytripping}, since the transition is modeled independent of the local condition of the free-stream which may differ from case to case. Furthermore, the key idea of  passive techniques is to trip the BL to re-energize the flow so that the flow remains attached~\cite{RANStrip}.
The skin friction, and consequently the drag, change due to the shift of the transition position from one test to another, which defines the turbulent portion of the model. The different transition onset in adverse-pressure-gradient (APG) flow can also result in larger separation regions farther downstream, with the corresponding impact on aerodynamic performance~\cite{nargesPhD}. In addition, it is sometimes possible to duplicate the relative thickness of the full-scale turbulent BL at certain locations on the wind-tunnel model by fixing transition at the proper location~\cite{Blackwell1969PreliminarySO}.

Most studies focusing on the physics of turbulent BLs employ tripping to promote early and robust transition to turbulence \cite{erm_joubert_1991,RamisAd,sanmigue}. For instance, different tripping devices were studied, in the context of a flat-plate BL~\cite{head_bandyopadhyay_1981,ExpTrip2017}. Roughness elements were also used to force transition in order to eliminate the transitional effects~\cite{ExpTrip2019,bypass}. The wide and continuous application of tripping in wind tunnels~\cite{recentTrp} motivated the use also in numerical methods to model similar effects in order to replicate the experimental data. Therefore, tripping methods evolved beside the numerical transition models that attempted to model the natural transition process. Referring to the variety of transition types and the complexity of modeling the complex physical interactions and mechanisms leading to transition, tripping models have the advantage of simplicity, which results in a much lower uncertainty. For instance, various tripping strategies were assessed over a flat plate by DNS~\cite{schlatter_orlu_2012} and were later implemented in large-eddy simulation (LES)  for a 2D airfoil~\cite{LEStrip}. 
\vspace{-6pt}
\section{Laminar--turbulent transition and tripping in EVM}
\label{EVM}
Among eddy-viscosity RANS models (EVM), the one-equation turbulence model of Spalart-–Allmaras (SA) contains a trip term~\cite{SA}. These authors used the word {\it trip} \say{to mean that the transition point is imposed by an actual trip, or natural but obtained from a separate method}~\cite{SA}. The trip version of the SA model, named as SA-Ia, is rarely used, because the model is most often employed for fully-turbulent (FT) applications~\cite{SAla}. Its trip term was found to be inadequate to force transition at a specified location, specifically for hypersonic flows~\cite{SAtrip}. The $k-\omega$ SST model, as the most common two-equations EVM model, was developed in 1994~\cite{SST_BSL}. The formulations of both turbulence transport equations ($k$ and $\omega$) are based on the features of FT-BL and so the initial laminar region, and consequently the transition part, are not modeled accurately. Such a formulation induces an early turbulent--viscosity buildup (as if for \emph{e.g.} there is a surface roughness) and therefore causes FT flow over the region which is laminar in the physical model and leads to over-estimating the drag~\cite{FoulingRoughness,nargesPhD}. 
RANS-based BL transition algorithms have been broadly considered in literature since few decades ago~\cite{kklmod,kklomega}. Most commonly transition models consist of two main parts: 1. Define the laminar, transition and FT regions, \emph{i.e.} the intermittency ($\gamma$) distribution. This can be done via two, one, or even zero transport equations~\cite{ansysGuidetheory,Menter,Langtry2006ACT}; 2. Apply the modifications into the turbulence model, \emph{i.e.} in the $k$ and $\omega$ equations, which is referred to as `coupling with $k-\omega$ SST' model. The currently available transition models in RANS are typically based on empirical correlations, but are not specifically aimed at representing the physical mechanisms in the transition process. As discussed by Langtry~\cite{Langtry2006ACT}, \say{They do not attempt to model the physics of the transition process (unlike \emph{e.g.}\ turbulence models), but form a framework for the implementation of transition correlations into general purpose CFD methods}. They are basically designed to cover the standard `bypass transition', as well as flows in low free-stream turbulence environments (since the transition location is correlated with the free-stream turbulence intensity, based on laboratory data)~\cite{ansysGuidetheory}. In addition to becoming unstable (in terms of convergence)~\cite{FoulingRoughness}, it was observed that setting a lower value for the free-stream turbulence in the CFD simulations would result in a later transition prediction than observed in the physical model~\cite{RANShyb}.
Apart from the pros and cons of such approaches in simulating the transition process, recent studies show that \say{there is potential for uncertainty or error in simulating the forced transition case with RANS models}\cite{FoulingRoughness}. In order to initiate transition at the same location as the experimental data, zigzag tapes were used as the model, but the uncertainties inevitably appeared even in the calculations of the integrated parameters, \emph{e.g.} the total power of the whole turbine rotor. Certainly, it is more challenging when a point-to-point comparison  is intended, \emph{e.g.} in chordwise $C_p$ distribution. Similarly, turbulence tripping was implemented in RANS using a specific type of obstacle in the geometry, which caused flow disturbances that facilitated the transition from laminar to turbulent flow~\cite{RANStrip}. Although a sudden jump in the local pressure was achieved, a spurious small vortex emerged downstream of the obstacle. 
According to section~2, tripping would be a required feature in RANS models, while on the numerical side, we found that there is no suitable model to implement tripping. In the following sections we discuss a method of tripping for k-$\omega$ SST.

\vspace{-6pt}
\section{Methodology}
\label{meth}
We start with describing a turbulence-generation mechanism, which was recently adopted by Fahland~\cite{Gerog} for the RANS simulation of the flow around an airfoil. We denote this as `injection method' ($k$I), and it is based on directly modifying the turbulent kinetic energy, $k$ at the target trip point. This technique serves as an efficient tripping and the results are in agreement with the other methods described in Ref.~\cite{tripping_3}.

Note that the injection of extra $k$ effectively promotes transition at the position or shortly downstream of it. 
The modelled transport equation for $k$ is may be written as  
\begin{equation}\label{kEq1}
\frac{\partial (\rho k)}{\partial t} +\frac{\partial (\rho u_j k)}{\partial x_j}= 
P_k-D_k+
\frac{\partial[(\mu+\sigma_k \mu_t)\frac{\partial k}{\partial x_j}] } {\partial x_j}+S_k \ ,
\end{equation}
where $P_k$ and $D_k$ denote the production and dissipation terms respectively~\cite{SST_BSL}.
The coefficients $\mu$ and ${\mu}_t $ denote the dynamic viscosity and turbulent dynamic viscosity, respectively, and the corresponding term (the third term on the right-hand side) refers to the diffusion of $k$. The last term on the right-hand side, $S_k$, is included to account for the sources of turbulent kinetic energy. The $k$I approach is based on adding a local $S_k$ at the target trip point so that the flow becomes FT immediately, in the same way as in the experimental tripping. Furthermore, the standard $k-\omega$ SST model leads to an early transition so that the BL becomes FT even before the physical transition position. In order to ensure the turbulence model does not lead to a premature deviation from the laminar solution, the value of $k$ can be set to zero in the domain just upstream the tripping. This constraint is set to avoid an over-estimation of $k$ in the laminar region, which is anyway calculated from the FT equations in the $k-\omega$ SST model.

\begin{figure}[h!]
     
     \begin{subfigure}{0.4\textwidth}
     
     \includegraphics[width=\textwidth]{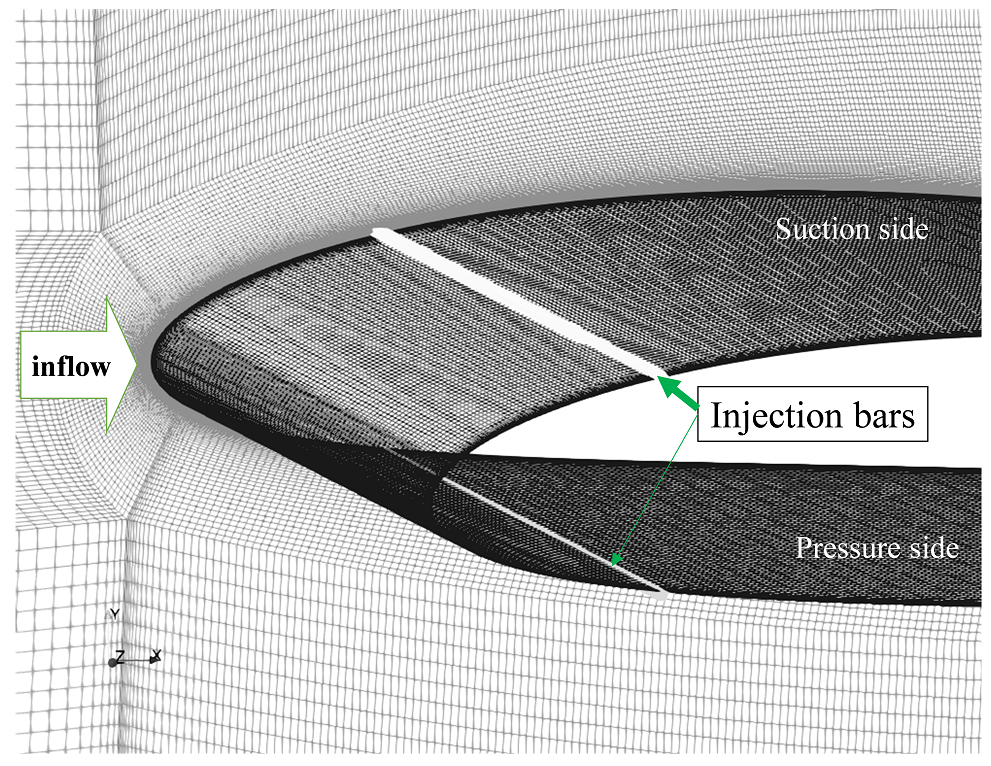}
    \label{fig:sketch1}\vspace{-16pt}
    \end{subfigure}
    \centering
    
     \begin{subfigure}{0.4\textwidth}
          \includegraphics[width=\textwidth]{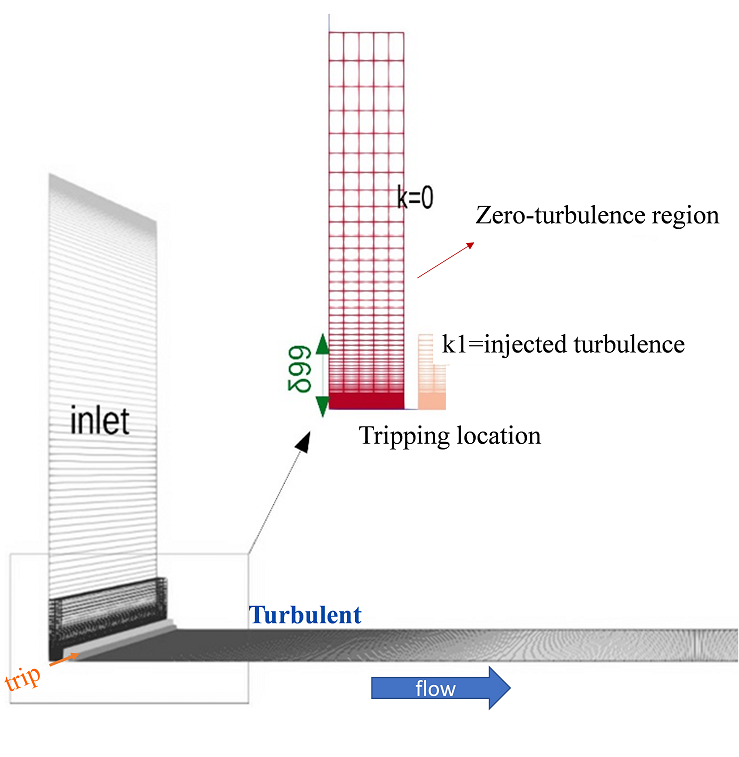}
        \label{fig:sketch2}
    \end{subfigure}
    \vspace{-16pt}
        \caption{Illustration of the injection--tripping procedure in various flows relevant for the present work. See text for more details.}
        \label{fig:1}
\centering
\end{figure}

Note that the value of $S_k$, the injection magnitude, should be large enough to raise the local skin-friction coefficient $C_f$ at the tripping point. An advantage of the $k$I method is the simplicity of the implementation, because the source terms can typically be externally modified  without changing the core of the solver. Fig.~\ref{fig:1} illustrates how the injection area is specified for a flat plate and a wing model, similar to tripping in the experiments. The laminar region, \emph{i.e.}\ the $k=0$ area, is defined to implement the turbulence constraint, which was defined at the beginning of this section. The approximated depth of the injected area is suggested to be approximately equal to the momentum thickness ${\theta}$ at the tripping location~\cite{Gerog}, since the tripping should be located inside the BL region. The $k_1$ part in Fig.~\ref{fig:1}~bottom shows the injection section, as well as the injection bar assigned on the wing in Fig.~\ref{fig:1}~top. 

The assessment includes three test cases. We consider the Minimum-Turbulence-Level (MTL) wind tunnel at KTH Royal Institute of Technology. The NACA4412 profile is the reference airfoil selected for this study. As the base turbulence model, a two-equation EVM model is considered: k-$\omega$ SST. OpenFOAM is used as the CFD solver.

\vspace{-6pt}
\section{Results}
\label{res}
The results are shown in terms of $C_f$, which is the normalized wall shear stress at the wall. The shape factor $H_{12}$ is also plotted as an indication of the boundary-layer development, since it is the ratio of the displacement to momentum thicknesses. For the mid-height section of the wing, the chordwise $C_p$ distribution is plotted, where the static pressure $p$ is non-dimensionalized with the dynamic pressure $P_d$. Three test cases are discussed in the following. Two main features are intended to ensure proper tripping: i) a sufficiently sharp $C_f$ increase, which is ii) immediate at the intended tripping location.

\begin{figure}[h!]
     \centering
     \hfill
     \centering
     \begin{subfigure}{.45\textwidth}
         \centering
         \includegraphics[width=\textwidth]{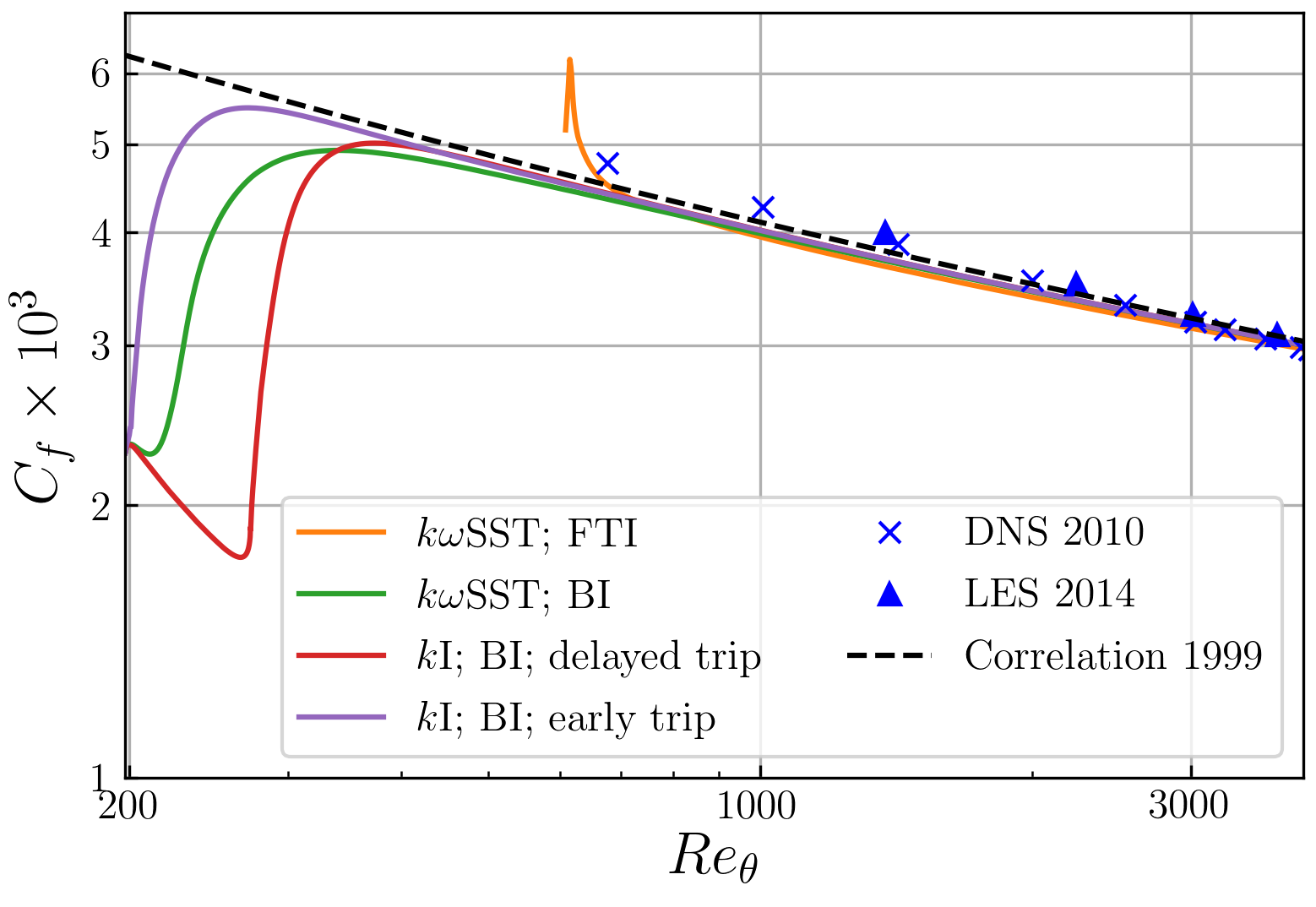}
         \caption{}
         \label{Bls_Cf}
     \end{subfigure}
     \hfill
     \vspace{-4pt}
     \begin{subfigure}{0.45\textwidth}
        \centering
         \includegraphics[width=\textwidth]{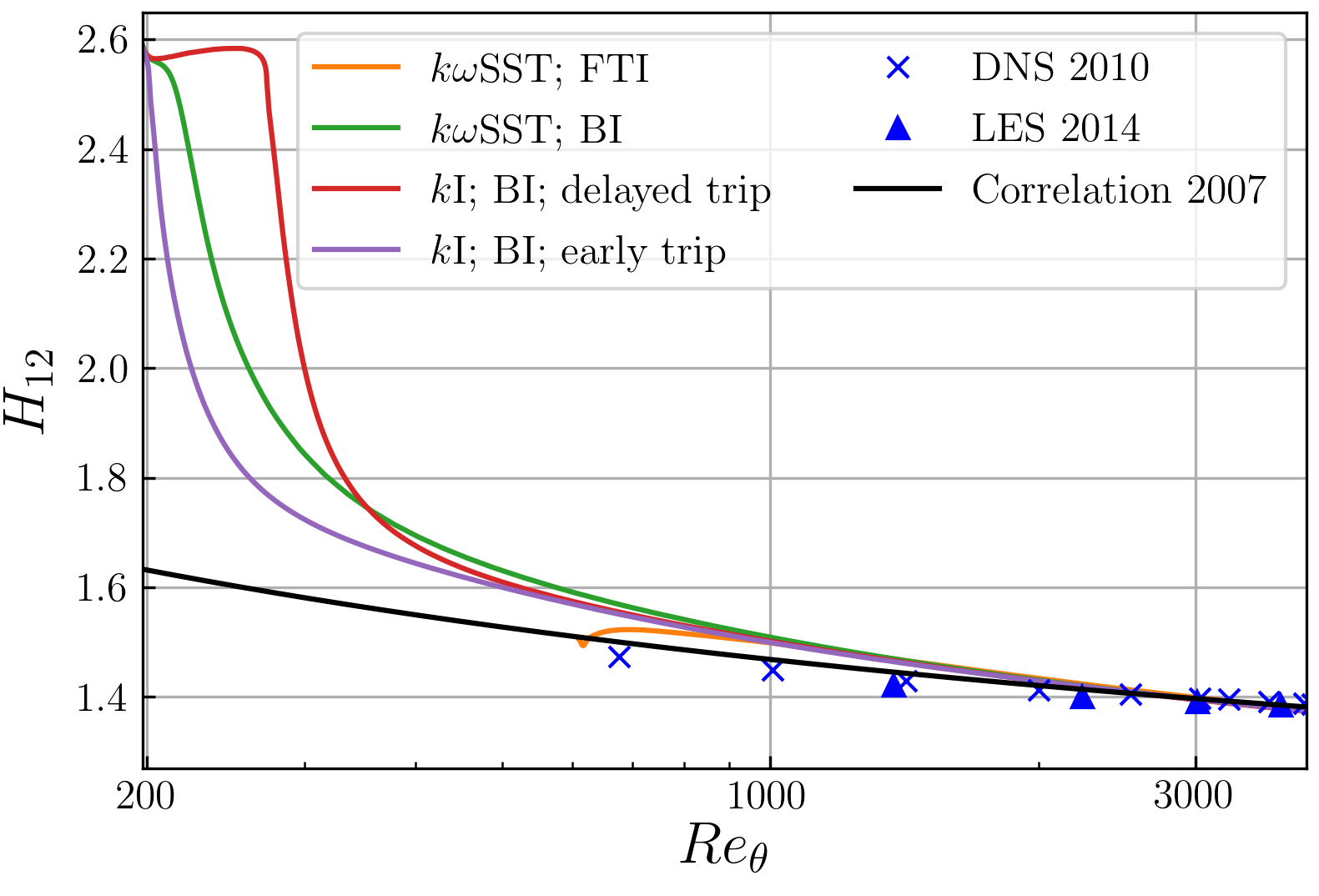}
         \caption{}
         \label{fig:Blas_H}
     \end{subfigure}
    \hfill
     \vspace{-10pt}
        \caption{Injection Tripping for ZPG flat plate, where we show the skin-friction coefficient $C_f$ and the shape factor $H_{12}$. Fully-turbulent baselines are based on Refs.~\cite{DNS_FT_2010,LES2014FT,SelfConsist_Monkewitz2007}; `FTI' denotes the fully-turbulent profile at inflow from DNS;`BI' denotes the Blasius inflow. $Re_{\theta}$ and $C_f$ are in log scale, while $H_{12}$ plot is in linear scale.} 
        \label{fig:ZPG}
\end{figure}

\textbf{I. Flat plate with zero pressure gradient (ZPG):}
In Fig.~\ref{fig:ZPG}, two scenarios are tested to assess the method efficiency: early and delayed tripping. First the flow is tripped immediately after the domain inlet as an `early-trip'. Conversely, in the case denoted as `delayed trip', the simulation keeps the flow laminar for some distance before it is tripped, see Ref.~\cite{tripping_3}.
It is shown that boundary-layer development can be controlled with this tripping method, which results in an adaptive laminar-turbulent transition.   

\begin{figure}[h!]
     \centering
         \includegraphics[width=.499\textwidth]{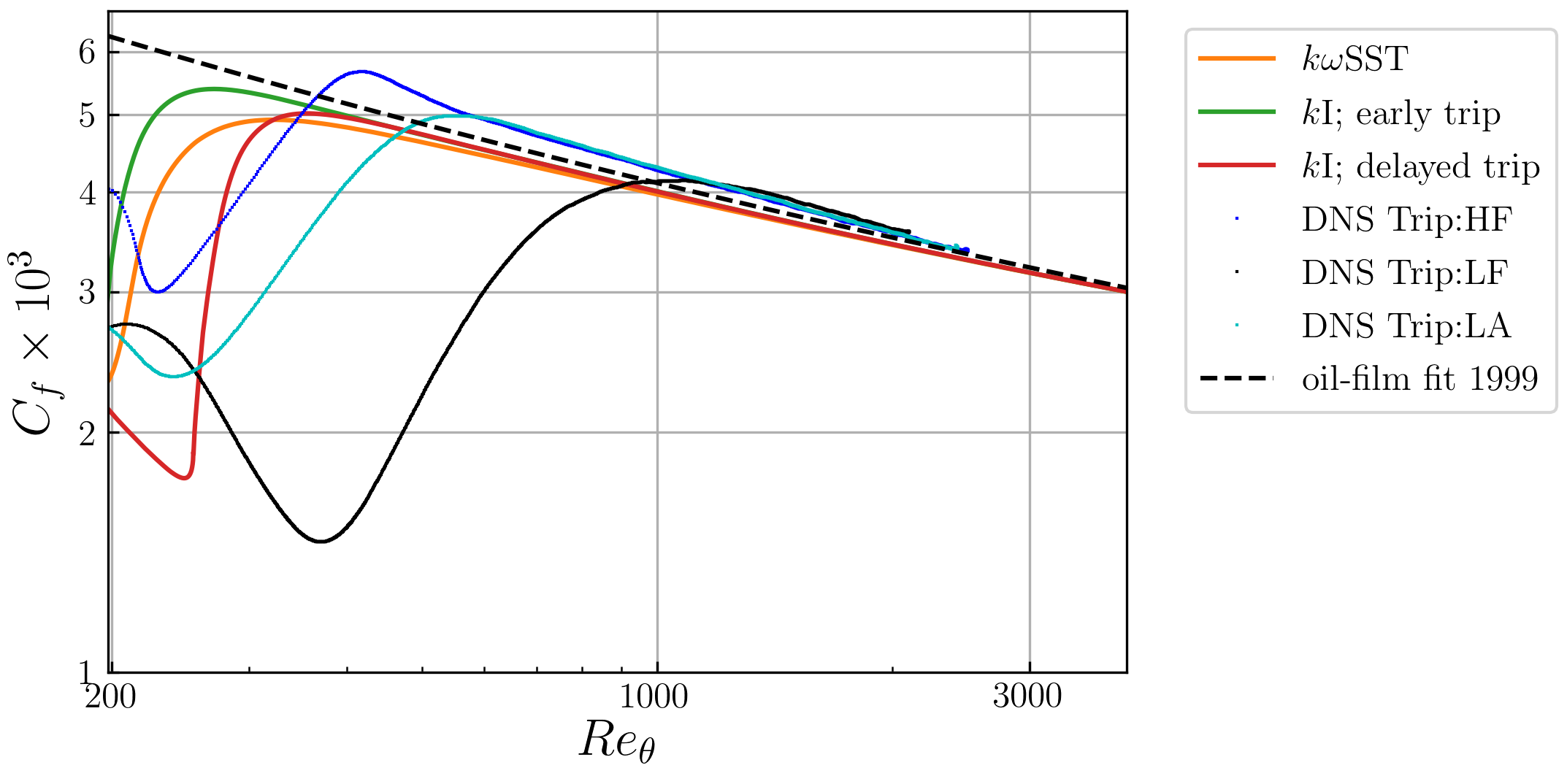}
     \hfill
 \vspace{-16pt}
 \caption{Comparison of the injection methods with various tripping parameters implemented in DNS~\cite{schlatter_orlu_2012}. FT baseline is according to Ref.~\cite{Osterlund8624} (oil-film fit 1999).}
\label{fig:DNS}
\end{figure}

To evaluate the $k$I tripping in RANS, in this part we focus on the low-$Re_{\theta}$ trends, which were studied in Ref.~\cite{schlatter_orlu_2012} via the use of different tripping parameters in DNS, see Fig.~\ref{fig:DNS}. The resulting turbulent flow with $k$I tripping quickly adapts to the canonical form of the turbulent BL, with shorter development length than the non-optimal tripping as studied by DNS.

\begin{figure}[h!]
\centering
     \begin{subfigure}{.45\textwidth}

        \includegraphics[width=\textwidth]{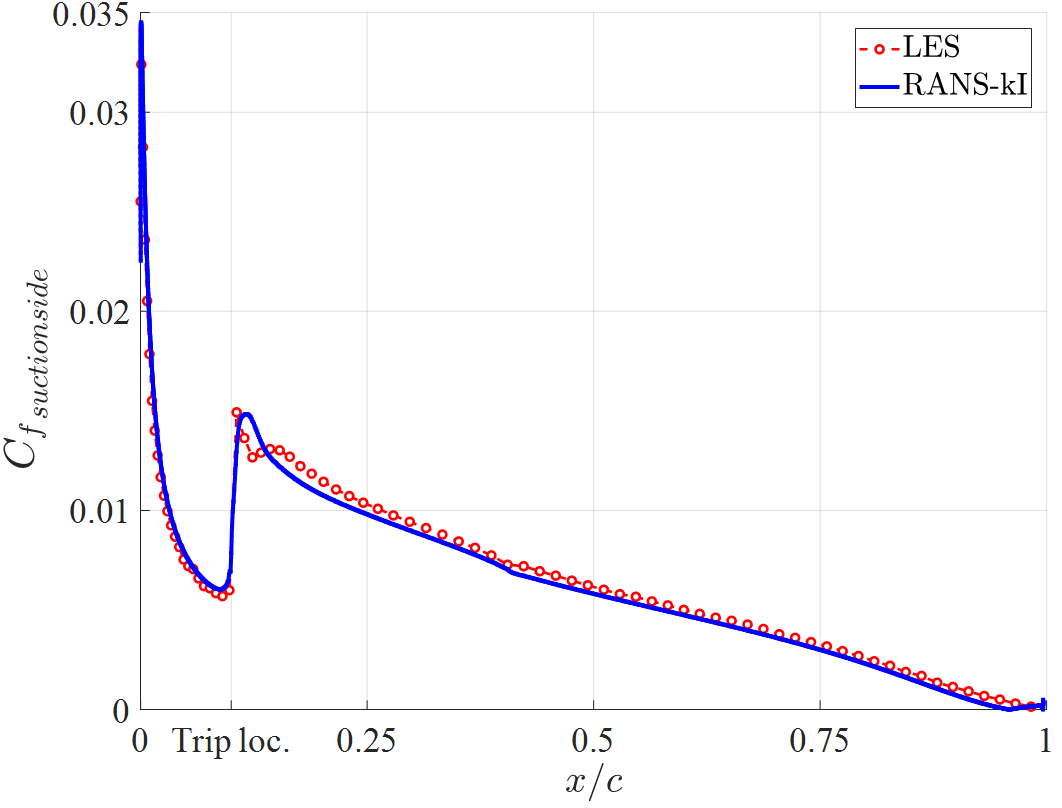}
        \caption{Injection Tripping: Well-resolved LES \cite{LES2014FT} vs RANS-$kI$ }
        \label{2D_tripp}
     \end{subfigure}

\begin{subfigure}{0.45\textwidth}

\includegraphics[width=\textwidth]{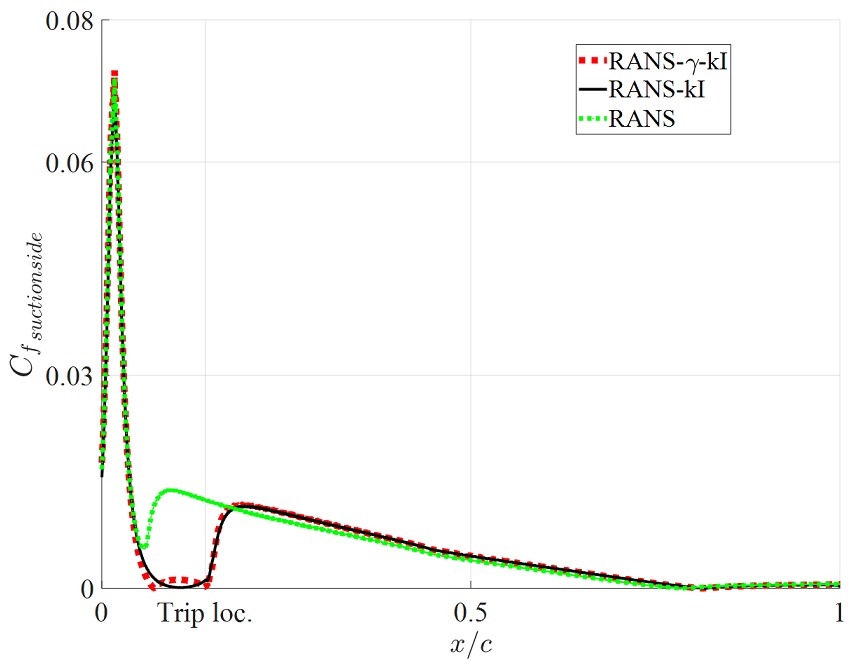}

     \caption{ Wind--tunnel tripping is compared to the reference data, denoted as RANS-$\gamma$-${k}$I in Ref.~\cite{tripping_3}.}
         \label{fig:inter}
\end{subfigure}

 \vspace{-4pt}
        \caption{Skin-Friction Factor $C_f$ for 2D cases at two angle of attacks (AOA): (a) isolated airfoil at AOA=$5^\circ$, (b) airfoil in a wind tunnel at AOA=$11^\circ$(b). }
        \label{2DWT}
\end{figure}

\textbf{II.	2D airfoil:} Tripping with RANS has been implemented for an isolated airfoil (a NACA4412 wing section in free-flight conditions) and compared with a well-resolved LES of the same case \cite{RicardADD} tripped with the method in Ref. \cite{schlatter_orlu_2012}. Skin-friction plots are in very good agreement, as observed in Figure \ref{2D_tripp}. The $k$I tripping (RANS-$k$I) is applied to an airfoil in a wind tunnel and the results are in agreement with another tripping method, described in Ref.~\cite{tripping_3}. The standard k-$\omega$ SST is denoted as RANS.


\begin{figure}[h!]
     \centering
     \begin{subfigure}{.49\textwidth}
         \centering
         \includegraphics[width=\textwidth]{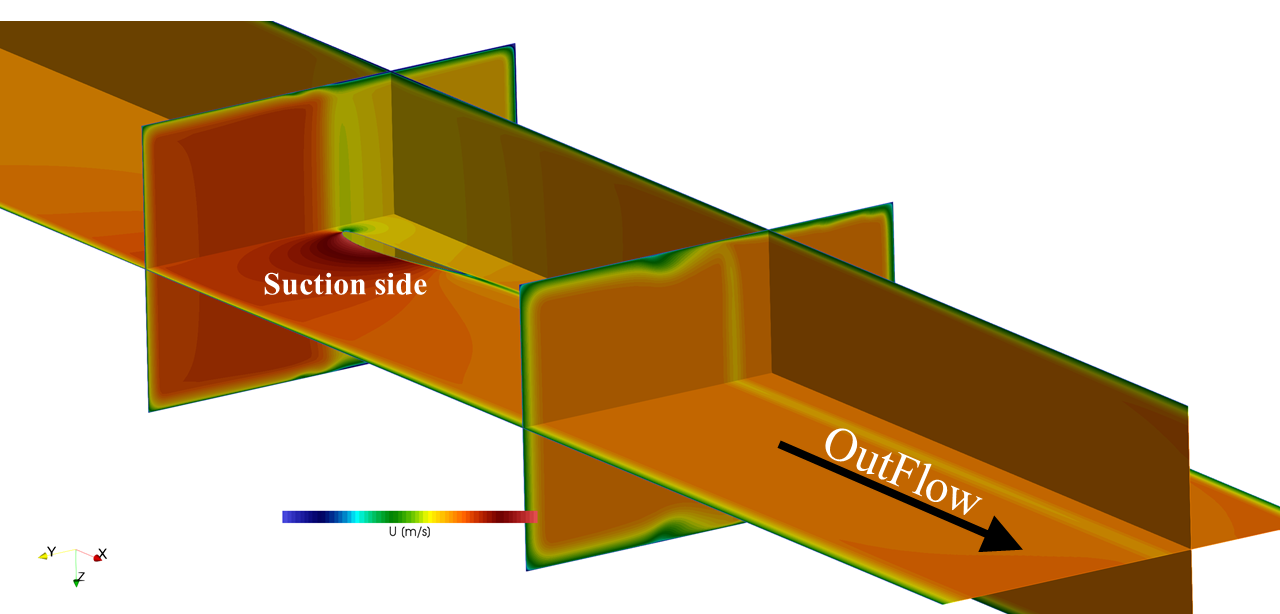}    
         \caption{Velocity-magnitude ontours}
         \label{fig:3D contour}
     \end{subfigure}
     \hfill
     \begin{subfigure}{0.45\textwidth}
        \centering
         \includegraphics[width=\textwidth]{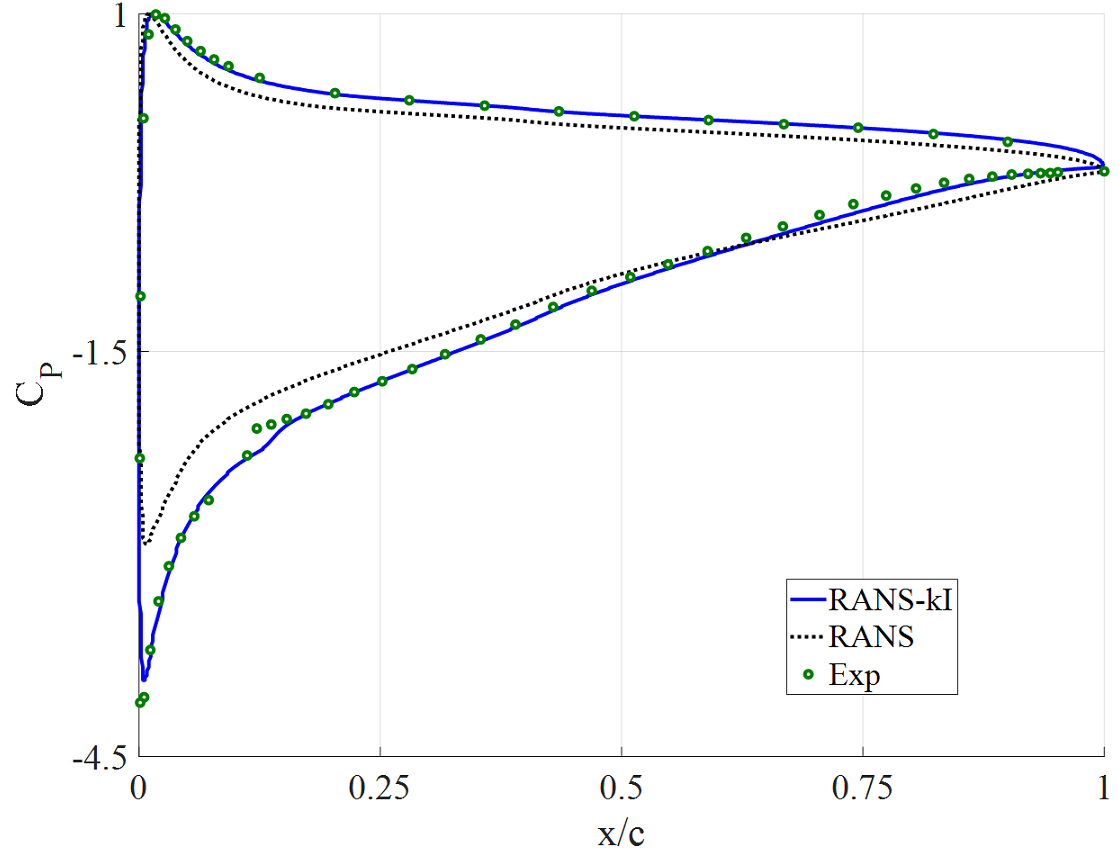}
         \caption{Pressure coefficient, $C_p$}
         \label{fig:cf3D}
     \end{subfigure}
     \hfill
 \vspace{-10pt}
        \caption{Tripping on a wing placed in a wind tunnel.}
\end{figure}

\textbf{III.	3D wing in a wind tunnel: }A 3D RANS simulation of wing at 11-degree angle of attack is performed considering the same tripping location as that in the wind-tunnel experiment.  The qualitative velocity contours at several selected sections are illustrated as well as the chordwise $C_p$ distribution (Figure \ref{fig:3D contour}). At such a high angle of attack, 3D RANS results in a lower suction compared to experiments, while a perfect agreement is achieved through the proposed tripping technique (Fig.~\ref{fig:cf3D}). Similar good agreement is also observed for $C_f$ (not shown here).
\vspace{-6pt}
\section{Conclusions and outlook}
An adaptive method for forced laminar--turbulent transition is assessed in this paper. Different applications of this tripping are discussed in an effort to replicate the tripped experimental tests, which is the specific purpose of this research. The implementation of laminar--turbulent tripping is assessed in a RANS--EVM  turbulence model with the purpose of developing more reliable aerodynamic simulations, in which the uncertain (and ultimately unnecessary) modeling of the transition process is avoided. Two main features are intended via the numerical tripping in the $k-{\omega}$SST model, which are according to the function of the experimental trip devices: first, the transition onset at the exact target trip location; and second a short development length. 
The results from the turbulence-injection ($k$I) method show a fair agreement with DNS and LES tripping approaches and experimental data. The tripping technique in 3D RANS simulation improves the results significantly so that a very good agreement between the experimental data and the 3D RANS is achieved. Therefore, the proposed tripping method is indicated as a potential approach to replicate experimentally--measured data from a real wind tunnel. This opens the way for faithful predictions of wind-tunnel experiments using RANS.



Financial support provided by the Knut and Alice Wallenberg Foundation is gratefully acknowledged. The computations were enabled by resources provided by the Swedish National Infrastructure for Computing (SNIC) at PDC and HPC2N.

%
\begin{References}
\printbibliography[heading=none]

@phdthesis{nargesPhD,
   author = {Tabatabaei, N.},
   institution = {Luleå University of Technology, Fluid and Experimental Mechanics},
   school = {Luleå University of Technology, Fluid and Experimental Mechanics},
   title = {Impact of Icing on Wind Turbines Aerodynamic},
   series = {Doctoral thesis / Luleå University of Technology 1 jan 1997 → …},
   ISSN = {1402-1544},
   ISBN = {978-91-7790-228-7},
   ISBN = {978-91-7790-229-4},
   year = {2018}
}

@article{RicardADD,
author = {Vinuesa, R. and Negi, P. and Atzori, M. and Hanifi, A. and Henningson, D. and Schlatter, P.},
year = {2018},
%month = {05},
pages = {86-99},
title = {Turbulent boundary layers around wing sections up to Re= 1,000,000},
volume = {72},
journal = {International Journal of Heat and Fluid Flow},
%doi = {10.1016/j.ijheatfluidflow.2018.04.017}
}

@article{sanmigue,
title={On the identification of well-behaved turbulent boundary layers},
volume={822}, 
%doi={10.1017/jfm.2017.258},
journal={Journal of Fluid Mechanics},
publisher={Cambridge University Press},
author={Sanmiguel Vila, C. and Vinuesa, R. and Discetti, S. and Ianiro, A. and Schlatter, P. and Örlü, R.}, year={2017}, pages={109–138}}

@article{schlatter_orlu_2012, 
title={Turbulent boundary layers at moderate Reynolds numbers: inflow length and tripping effects}, 
volume={710}, %doi={10.1017/jfm.2012.324},
journal={Journal of Fluid Mechanics},
publisher={Cambridge University Press}, 
author={Schlatter, P. and Örlü, R.},
year={2012},
pages={5–34}
}

@mastersthesis{Gerog,
  title={Flow Control for Turbulent Skin-Friction Drag Reduction on Airfoils},
  author={G. Fahland},
  institution = {Karlsruhe Institute of Technology},
    
  school = {Institute of Fluid Mechanics},
  year={2019}
}

@phdthesis{Langtry2006ACT,
  title={A correlation-based transition model using local variables for unstructured parallelized CFD codes},
  author={R. Langtry},
   institution = {University of Stuttgart},
  year={2006}
}

@article{insert_1,
author = {Tabatabaei, N. and Örlü, R. and Vinuesa, R. and Schlatter, P.},

title = {Aerodynamic free-flight conditions in wind-tunnel modelling through reduced-order wall inserts},

JOURNAL = {Fluids (Under review)},
VOLUME = {14},
YEAR = {2021},
}

@article{tripping_3,
author = {Tabatabaei, N. and Vinuesa, R. and Örlü, R. and Schlatter, P.},
year = {2021},
title = {New tripping method for RANS simulations of complex turbulent boundary layers},
journal = {Flow, Turbulence and Combustion (Under review)},
}

@article{SelfConsist_Monkewitz2007,
author = {Monkewitz, P.A.  and Chauhan, K.A.  and Nagib, H.M. },
title = {Self-consistent {high-Reynolds}-number asymptotics for zero-pressure-gradient turbulent boundary layers},
journal = {Physics of Fluids},
volume = {19},
number = {11},
pages = {115101},
year = {2007},
}

@phdthesis{Osterlund8624,
   author = {{\"O}sterlund, J.M.},
   institution = {KTH, Mechanics},
   note = {NR 20140805},
   pages = {xiii, 26},
   publisher = {Mekanik},
   school = {KTH, Mechanics},
   title = {Experimental studies of zero pressure-gradient turbulent boundary layer flow},
   series = {Trita-MEK},
   number = {99:10},
   year = {1999},
}

@article{Menter,
author = {Menter, F. and Smirnov, P. and Liu, T. and Avancha, R.},
year = {2015},
month = {07},
pages = {1-37},
title = {A One-Equation Local Correlation-Based Transition Model},
volume = {95},
journal = {Flow, Turbulence and Combustion},
}

@article{SST_BSL,
author = {Menter, F.R.},
title = {Two-equation eddy-viscosity turbulence models for engineering applications},
journal = {AIAA Journal},
volume = {32},
number = {8},
pages = {1598-1605},
year = {1994},
}

@manual{ansysGuidetheory,
    author    = "ANSYS, Inc.",
    title     = "ANSYS CFX-Solver Theory Guide",
    year      = "2015",
    publisher = "ANSYS, Inc.",
note= "Release 16.2."
   
}

@article{LES2014FT,
title = "Simulation and validation of a spatially evolving turbulent boundary layer up to {Re}=8300",
journal = "International Journal of Heat and Fluid Flow",
volume = "47",
pages = "57 - 69",
year = "2014",
author = "Eitel-Amor, G. and R. Örlü and P. Schlatter",
}

@article{DNS_FT_2010,
title={Assessment of direct numerical simulation data of turbulent boundary layers}, volume={659}, 
journal={Journal of Fluid Mechanics}, publisher={Cambridge University Press}, 
author={Schlatter,P.  and \"Orl\"u, R.}, year={2010}, pages={116–126}}

@article{RANStrip,
title = "Numerical study on effect of boundary layer trips on aerodynamic performance of E216 airfoil",
journal = "Engineering Science and Technology, an International Journal",
volume = "21",
number = "1",
pages = "77 - 88",
year = "2018",
author = "B.K. Sreejith and A. Sathyabhama",
}

@inproceedings{RANShyb,
author = {Luckring, J. and Deere, K. and Childs, R. and Stremel, P. and Long, K.},
booktitle = {32nd AIAA Aerodynamic Measurement Technology and Ground Testing Conference},
year = {2016},
month = {06},
pages = {},
title = {An Application of {CFD} to Guide Forced Boundary-Layer Transition for Low-Speed Tests of a Hybrid Wing-Body Configuration},
}

@mastersthesis{LEStrip,
  title={Effect of Boundary-Layer Tripping on Turbulence Generation and Trailing-Edge Noise in Transitional Airfoils},
  author={J.B. Lewis},
   institution = {Karlsruhe Institute of Technology},
    
   school = {Embry-Riddle Aeronautical University},
  year={2017}
}

@article{ExpTrip2017,

	year = 2017,
    month = {4},
	publisher = {{IOP} Publishing},
	volume = {822},
	pages = {012016},
	author = {K. Rengasamy and A.Ch. Mandal},
	title = {Experiments on effective tripping device in a zero pressure gradient turbulent boundary layer},
	journal = {Journal of Physics: Conference Series},
}

@article{ExpTrip2019,
title = "Boundary layer tripping on moderate Reynolds number oscillating foils",
journal = "Journal of Fluids and Structures",
volume = "86",
pages = "1 - 12",
year = "2019",
author = "D. Iverson and M. Boudreau and G. Dumas and P. Oshkai",

}

@article{erm_joubert_1991, title={Low-{Reynolds}-number turbulent boundary layers}, volume={230},
journal={Journal of Fluid Mechanics}, 
publisher={Cambridge University Press},
author={Erm, L.P. and Joubert, P.N.},
year={1991}, 
pages={1–44}
}

@article{RamisAd,
author = {Marusic, I. and Chauhan, K. and Kulandaivelu, V. and Hutchins, N.},
year = {2015},
%month = {11},
pages = {379-411},
title = {Evolution of zero-pressure-gradient boundary layers from different tripping conditions},
volume = {783},
journal = {Journal of Fluid Mechanics},

}

@article{kklomega,
  title={A Three-Equation Eddy-Viscosity Model for Reynolds-Averaged Navier-Stokes Simulations of Transitional Flow},
  author={D.K. Walters and D. Cokljat},
  journal={Journal of Fluids Engineering-transactions of The Asme},
  year={2008},
  volume={130},
  pages={121401}
}

@article{SAtrip,
author = {Roy, Ch. and Blottner, F.},
year = {2001},
month = {09},
pages = {699-710},
title = {Assessment of One- and Two-Equation Turbulence Models for Hypersonic Transitional Flows},
volume = {38},
journal = {Journal of Spacecraft and Rockets},
%doi = {10.2514/2.3755}
}

@article{FoulingRoughness,
author = {Acarer,S.},
title = {Critical study of the effects and numerical simulations of boundary layer transition in lift-based wind turbines at moderate Reynolds numbers},
journal = {Journal of Renewable and Sustainable Energy},
volume = {12},
number = {6},
pages = {063309},
year = {2020},
%doi = {10.1063/5.0020357},

%URL = {         https://doi.org/10.1063/5.0020357
    
},
%eprint = {        https://doi.org/10.1063/5.0020357
    
}

}

@article{bypass,
author = {Reshotko, E. },
title = {Transient growth: A factor in bypass transition},
journal = {Physics of Fluids},
volume = {13},
number = {5},
pages = {1067-1075},
year = {2001},
}

@MISC{SAla,
author = {Rumsey, Ch. },
title = {Turbulence Modeling Resource},
month = Jan,
year = {2021},
note = {Langley Research Center},
}

@article{scale,
title = {A review of scale effects in unsteady aerodynamics},
journal = {Progress in Aerospace Sciences},
volume = {28},
number = {4},
pages = {273-321},
year = {1991},
author = {D.G. Mabey},
}

@techreport{whytripping,
  title={Boundary-layer velocity profiles downstream of three-dimensional transition trips on a flat plate at {M}ach 3 and 4},
  author={J.B. Peterson},
  year={1969},
  institution={NASA Langley Research Center Hampton, VA, United States}
}

@techreport{Blackwell1969PreliminarySO,
  title={Preliminary study of effects of Reynolds number and boundary-layer transition location on shock-induced separation},
  author={J.A. Blackwell},
  year={1969},
  institution={NASA Langley Research Center Hampton, VA, United States}
}

@inbook{recentTrp,
author = {F. Leticia dos Santos and M.P. Sanders and L. Dantas de Santana and C.H. Venner},
title = {Influence of tripping devices in hastening transition in a flat plate submitted to zero and favorable pressure gradients},
booktitle = {AIAA Scitech 2020 Forum},
chapter = {},
pages = {},

}

@inproceedings{SA,
  title={A One-Equation Turbulence Model for Aerodynamic Flows},
  author={P. Spalart},
  year={1992},
  booktitle = {30th Aerospace Sciences Meeting and Exhibit},
}

@phdthesis{kklmod,
author = {Gokdepe, M.},
year = {2015},
month = {01},
title = {Turbulence models for the numerical prediction of transitional flows with RANSE},
institution = {Massachusetts Institute of Technology},
school = { Department of Mechanical Engineering}
}

@article{head_bandyopadhyay_1981, title={New aspects of turbulent boundary-layer structure}, volume={107}, 
%DOI={10.1017/S0022112081001791}, 
journal={Journal of Fluid Mechanics}, publisher={Cambridge University Press}, 
author={Head, M.R. and Bandyopadhyay, P.}, year={1981}, pages={297–338}}
\end{References}
\end{document}